\documentclass[showpacs,amssymb,aps,preprint]{revtex4}
\usepackage{amsmath}
\usepackage{amstext}
\usepackage{amsopn}
\usepackage{amsfonts}
\usepackage{amssymb}
\usepackage{bbm}
\usepackage{accents}
\usepackage{empheq}
\usepackage{graphicx}
\usepackage{epsf}
\usepackage{graphics}
\usepackage[latin1]{inputenc}
\def\sc{\scriptscriptstyle}

\begin{document}

\title{An alternative construction of the positive inner product in non-Hermitian quantum mechanics}

\author{Ashok Das$^{a,b}$ and L. Greenwood$^{a}$\footnote{$\ $ e-mail: das@pas.rochester.edu,  lgreenwo@pas.rochester.edu}}
\affiliation{$^a$ Department of Physics and Astronomy, University of Rochester, Rochester, NY 14627-0171, USA}
\affiliation{$^b$ Saha Institute of Nuclear Physics, 1/AF Bidhannagar, Calcutta 700064, India}

\begin{abstract}
Within the context of non-Hermitian quantum mechanics, we use the generators of eigenvectors of the Hamiltonian to construct a unitary inner product space.  Such models have been of interest in recent years, for instance, in the context of ${\cal PT}$ symmetry, although our construction extends to the larger class of so-called pseudo-Hermitian Operators.  We provide a detailed example to illustrate the concept and compare with known results.  
\end{abstract}


\maketitle
\newpage
Let us consider a non-Hermitian quantum mechanical Hamiltonian operator, $H$, which is related to its adjoint through a similarity transformation
\begin{equation}
H=S^{-1}H^\dagger S.\label{similarity}
\end{equation}
An operator with the above relation is sometimes called pseudo-Hermitian \cite{mo}.  The adjoint, $H^\dagger$, is defined with respect to an auxilary Hilbert space, $\mathcal{H}$, equipped with an inner product $\langle\cdot|\cdot\rangle$ (conventionally taken to be the Dirac inner product), such that $\langle \phi|H\psi\rangle=\langle H^\dagger\phi|\psi\rangle$, with $|\psi\rangle,\ |\phi\rangle\in\mathcal{H}$. Relation \eqref{similarity} implies that $S$ can be chosen to be self-adjoint.

Since $H$ is not Hermitian, it follows that $e^{-itH}$  (for $t\in\mathbb{R}, \hbar =1$) is not unitary and, therefore, it is not possible to construct a unitary quantum theory on $\mathcal{H}$.  To construct a unitary quantum theory  we must define a new Hilbert space, $\mathcal{H}'$, with a modified inner product leading to a modified adjoint,  $H^{\ddagger}$, for which the Hamiltonian $H$ is self-adjoint.  By choosing $\mathcal{H}'$ so that $H^{\ddagger}=H$, we can guarantee that $e^{-itH}$ is a unitary time evolution operator in $\mathcal{H}'$.

Before we begin, let us make the following observation on the quadratic form in $\mathcal{H}$ defined as 
\begin{eqnarray}
\langle \phi |\psi\rangle_{\sc S}:=\langle\phi|S|\psi\rangle.\label{sproduct}
\end{eqnarray}
The adjoint with respect to this quadratic form is given by 
\begin{eqnarray}
\langle \phi |H\psi\rangle_{\sc S}=\langle H^{\#}\phi|\psi\rangle_{\sc S},
\end{eqnarray}
where,
\begin{eqnarray}
H^{\#}=S^{-1}H^\dagger S.
\end{eqnarray}
It follows from \eqref{similarity} that 
\begin{equation}
H^{\#}=H,
\end{equation} 
so that $H$ is self-adjoint, and time evolution is unitary with respect to this quadratic form.  However, if the eigenstates of the Hamiltonian are given by 
\begin{equation}
H|\psi_{\sc E}\rangle=E|\psi_{\sc E}\rangle,\quad E\in \mathbbm{C},\label{energystates}
\end{equation}
then, it follows that 
\begin{eqnarray}
(E-\bar{E}')\langle\psi_{{\sc E}'}|\psi_{\sc E}\rangle_{\sc S}=\langle\psi_{{\sc E}'}|S(H-H)|\psi_{\sc E}\rangle=0,\label{orthonormality}
\end{eqnarray}
where $\bar{E}$ denotes the complex conjugate of $E$. Equation \eqref{orthonormality} implies that 
\begin{eqnarray}
\langle\psi_{{\sc E}'}|\psi_{\sc E}\rangle_{\sc S}=e^{-iF(E)}\delta_{\sc E\bar E'},\label{orthonormality1}
\end{eqnarray}
where the constant $e^{-iF(E)}$ depends only on $E$ because of the delta function, and may be negative (for instance, if $S$ has negative eigenvalues). Consequently, while the quadratic form \eqref{sproduct} leads to a unitary time evolution, it cannot be considered an inner product. 
 
We can turn this quadratic form into an inner product if we can remove the phase $e^{-iF(E)}$ in \eqref{orthonormality1}, which will ensure that $\langle\psi_{\bar{\sc E}}|\psi_{\sc E}\rangle_{\sc S}\geq0$ for all $|\psi_{\sc E}\rangle\in\mathcal{H}'$.  If $S$ is a positive operator, so that it can be written as $S=g^\dagger g$, then the quadratic form \eqref{sproduct} is positive definite and, therefore, an inner product. (This will be the case, for example, when $H$ is related to a Hermitian Hamiltonian $h$ through a similarity transformation $H = g^{-1}hg, h^{\dagger}=h$.)  We will consider the case where $S$ is not positive.  We propose that there exists an operator $A$, which commutes with $H$ such that \begin{equation}
q=SA,\label{q}
\end{equation} 
is positive on a suitably defined Hilbert space, $\mathcal{H}'$,
\begin{equation}
\langle \phi |\phi\rangle_q:=\langle\phi|q|\phi\rangle = \langle\phi|SA|\phi\rangle\geq0,\quad {\rm for\ all\ }|\phi\rangle\in\mathcal{H}'.\label{positivity}
\end{equation}
Since  $[A,H]=0$, we have 
\begin{eqnarray}
H^\ddagger:=q^{-1}H^\dagger q=A^{-1}S^{-1}H^\dagger S A = A^{-1} H A = H,\label{hddagger}
\end{eqnarray}
where $H^\ddagger$ denotes the adjoint with resect to $\langle\cdot|\cdot\rangle_q$.  

We define the spectrum of $H$ to be 
\begin{eqnarray}
{\rm spect}(H)=\{E\ \ {\rm s.t.}\  0<|\langle\psi_{\bar{\sc E}}|\psi_{\sc E}\rangle_S|<\infty\},
\end{eqnarray}
and note that \eqref{similarity} implies  $H$ and $H^\dagger$ are isosepctral.
Let the projection operator $P_{\sc E}$, for $E\in {\rm spect}(H)$, satisfy the following identities
\begin{eqnarray}
P_{\sc E}|\psi_{\sc E}\rangle&=&|\psi_{\sc E}\rangle,\nonumber\\
P_{\sc E}|\psi_{{\sc E}'}\rangle&=&0,\quad E\neq E',
\end{eqnarray}
and furthermore, let
\begin{equation}
P_{{\sc\mathcal{H}}'}=\sum_{{\sc E}\in\;{\rm spect}(H)}P_{\sc E}.
\end{equation}
We will define $\mathcal{H}'=$Im $P_{{\sc\mathcal{H}}'}$ (the image of $P_{{\sc\mathcal{H}}'}$) with,
\begin{equation}
 \mathbbm{1}_{{\sc\mathcal{H}}'}=P_{{\sc\mathcal{H}}'}.\label{completeness}
\end{equation}  
$\mathcal{H}'$ equipped with $\langle\cdot|\cdot\rangle_{q}$ becomes an inner product space once we define $A$ which makes $q$ positive.  It follows from \eqref{hddagger} that $e^{-itH}$ is also unitary for real $t$. 

We note that by construction \{$|\psi_{\sc E}\rangle$\} is complete in $\mathcal{H'}$ and, therefore, any operator $A$ that commutes with $H$ can be expressed as \cite{reed},
\begin{eqnarray}
A=\sum_{\sc E} c_{\sc E}\, P_{\sc E},\label{A}
\end{eqnarray}
where $c_{\sc E}\in\mathbb{C}$, and it is understood that $E\in{\rm spect}(H)$.  It follows that 
\begin{eqnarray}
q=S\sum_{\sc E} c_{\sc E}\, P_{\sc E},
\end{eqnarray}
which, upon using \eqref{orthonormality1},  leads to,
\begin{equation}
\langle\psi_{{\sc E}'}|\psi_{\sc E}\rangle_q=\langle\psi_{{\sc E}'}|S\sum_{{\sc E}''} c_{{\sc E}''}\ P_{{\sc E}''}|\psi_{\sc E}\rangle= c_{\sc E} \langle\psi_{{\sc E}'}|S|\psi_{\sc E}\rangle = c_{\sc E}\;e^{-iF(E)}\delta_{E\bar E'}.\label{qproduct}
\end{equation}
In order to ensure that the right hand side is positive for all $|\psi_{\sc E}\rangle\in\mathcal{H'}$ we choose 
\begin{equation}
c_{\sc E}=e^{ iF(E)}=\langle\psi_{\bar{\sc E}}|\psi_{\sc E}\rangle^{-1}_{\sc S},
\end{equation}
thereby defining the action of $A$.  It follows that 
\begin{equation}
q = S\sum_{\sc E} e^{ iF(E)}\ P_{\sc E} = Se^{ iF(H)}\sum_{\sc E} \ P_{\sc E} = Se^{ iF(H)} \mathbbm{1}_{{\sc\mathcal{H}}'},\label{qq}
\end{equation}
which gives 
\begin{equation}
\langle\psi_{{\sc E}'}|\psi_{\sc E}\rangle_q=\delta_{\sc E\bar E'},\quad {\rm for\ all\ }|\psi_{\sc E}\rangle\in\mathcal{H'}.
\end{equation}

Comparing \eqref{q} with \eqref{qq} (and using \eqref{completeness}), we determine that  $A=e^{iF(H)}=\langle\psi_{\bar{\sc E}}|\psi_{\sc E}\rangle_{\sc S}^{-1}$ ($E\rightarrow H$ with $H$ on the right), whenever $\langle\psi_{\bar{\sc E}}|\psi_{\sc E}\rangle_{\sc S}$ is a smooth function of $E$.  However, the substitution $E\rightarrow H$ may not be straightforward to carry out in practice, perhaps due to discontinuities in $\langle\psi_{\bar{\sc E}}|\psi_{\sc E}\rangle_{\sc S}$ (or as we shall see in the example, if the Hamiltonian describes more than one particle).  For this reason we describe an alternative method of solving for $q$, assuming that one has already determined the operator $\sigma_{\sc E}$ which generates the eigenvector $|\psi_{\sc E}\rangle$.

Let us consider the following operator equation which defines $\sigma_{\sc E}$:
\begin{equation}
H\sigma_{\sc E}=E\sigma_{\sc E}+\sigma_{\sc E}k_{\sc E}.\label{Hsigma}
\end{equation}
We shall call $\sigma_{\sc E}$ a ``generator'' for the eigenstates of $H$ if the following three conditions are satisfied$^{\footnotemark[1]}$ \footnotetext[1] {$\ $ If $|\psi_{\sc E}\rangle$ is $i$-fold degenerate we require a $\sigma^i_{\sc E}$ for each degenerate state.}.

\begin{description}
\item[\textmd{$(i)$ }] There exists at least one vector $|\psi\rangle$, solving
\begin{equation}
k_{\sc E}|\psi\rangle=0,\quad {\rm for\ all\ }E\in{\rm spect}(H),
\end{equation}
with $\sigma_{\sc E}|\psi\rangle\neq0$.
\item\textmd{$(ii)$ } There exists at least one vector $|\phi\rangle$ solving
\begin{equation}
k_{\sc E}^\dagger|\phi\rangle=0,\quad  {\rm for\ all\ }E\in{\rm spect}(H),\label{kdagger}
\end{equation}
with $\sigma_{\sc E}^{\dagger-1}|\phi\rangle\neq0$, and $\langle\psi|\phi\rangle\neq0$.  
\item\textmd{$(iii)$ } $\sigma_{\sc E}$ has an inverse $\sigma_{\sc E}^{-1}$, at least on the subspace $P_{\sc E}$, and an adjoint-inverse $\sigma_{\sc E}^{\dagger-1}$, at least on $|\phi\rangle$ defined in \eqref{kdagger}.
\end{description}

\noindent For instance, if one has solved for $\tau$ which satisfies the commutation relation $[\tau,H]=i$, then $\sigma_{\sc E}=e^{iE \tau}$ satisfies the three  conditions with $k_{\sc E}=(H - \lambda \mathbbm{1})$, corresponding to the reference state $|\psi\rangle=|\psi_\lambda\rangle$, the state with energy eigenvalue $\lambda$.  In general, we can expect $\sigma_{\sc E}$ to converge only for values of $E\in{\rm spect}(H)$ \cite{time}.

Condition $(i)$ implies that 
\begin{equation}
|\psi_{\sc E}\rangle=\sigma_{\sc E}|\psi\rangle,\label{psiE}
\end{equation}
is an eigenvector of $H$ with eigenvalue $E$. Furthermore, taking the adjoint of \eqref{Hsigma} and formally acting on both sides with $\sigma_{\sc E}^{\dagger-1}$ we get
\begin{eqnarray}
H^{\dagger}\;\sigma_{\sc E}^{\dagger-1}=\bar{E}\;\sigma_{\sc E}^{\dagger-1}+\sigma_{\sc E}^{\dagger-1}\;k_{\sc E}^\dagger.\label{Hsigmadagger}
\end{eqnarray}
It follows from condition $(ii)$ that  
\begin{equation}
|\phi_{\sc E}\rangle=\sigma_{\sc E}^{\dagger-1}|\phi\rangle,\label{phiEbar}
\end{equation} 
is an eigenvector of $H^\dagger$ with eigenvalue $\bar{E}$.  

Finally, from \eqref{psiE} and \eqref{phiEbar} we note that the eigenvectors of $H^\dagger$ with eigenvalue $E$ can be written either as $S|\psi_{\sc E}\rangle = S \sigma_{\sc E}|\psi\rangle$, or as $\sigma_{\bar{\sc E}}^{\dagger-1}|\phi\rangle$.  Therefore, these expressions must differ by a multiplicative  constant depending only on $E$,
\begin{eqnarray}
c'_{\sc E} S|\psi_{\sc E}\rangle=\sigma_{\bar{\sc E}}^{\dagger-1}|\phi\rangle.\label{cprime0}
\end{eqnarray}
We propose that 
\begin{eqnarray}
c'_{\sc E}=c\;e^{iF(E)},\label{cprime}
\end{eqnarray}
where $c$ is a constant independent of $E$ and which can be seen as follows. Indeed, acting on both sides of \eqref{cprime0} with $\langle\psi_{\bar{\sc E}}|$, and using \eqref{orthonormality1} as well as \eqref{psiE} we obtain
\begin{equation}
c'_{\sc E}e^{-iF(E)} = \langle\psi_{\bar{\sc E}}|\sigma_{\bar{\sc E}}^{\dagger-1}|\phi\rangle = \langle\psi|\sigma_{\bar{\sc E}}^\dagger\sigma_{\bar{\sc E}}^{\dagger-1}|\phi\rangle = \langle\psi|\phi\rangle.
\end{equation}
By conditions ($i$) and ($ii$), the right hand side is independent of $E$, so we get $c'_{\sc E}=\langle\psi|\phi\rangle\,e^{iF(E)} = c\, e^{iF(E)}$.

We therefore define the action of $q$ on the eigenstates of $H$ as 
\begin{eqnarray}
q|\psi_{\sc E}\rangle=q\sigma_{\sc E}|\psi\rangle:=\sigma_{\bar{\sc E}}^{\dagger-1}|\phi\rangle,
\end{eqnarray}
This is our main result.  If we let $q_0:|\psi\rangle\rightarrow|\phi\rangle$ (for instance, $q_0$ might equal $S$) and normalize $\langle\psi|\phi\rangle=1$, then we can express $q$ as,
\begin{equation}
q=\sum_{\sc E} \sigma_{\bar{\sc E}}^{\dagger-1}\;q_0\;\sigma_{ E}^{-1}P_{\sc E},
\end{equation}
and this leads to
\begin{equation}
\langle\psi_{{\sc E}'}|\psi_{\sc E}\rangle_q = \langle\psi_{{\sc E}'}|q|\psi_{\sc E}\rangle = \langle\psi_{{\sc E}'}|\sigma_{\bar{\sc E}}^{\dagger-1}|\phi\rangle = \langle\psi|\sigma_{{\sc E}'}^\dagger\sigma_{\bar{\sc E}}^{\dagger-1}|\phi\rangle = \delta_{\sc E\bar E'}.
\end{equation}
It is worth emphasizing here that this construction of $q$ applies equally well in the case  when the energy eigenvalues are complex as when they are real. As in the previous case, we can make the replacement $E\rightarrow H$ (with the $H$'s to the right) if it is possible to express $q$ in a power series in $E$.  This procedure is demonstrated in the following example.

We note that the strongest restriction on $\sigma_{\sc E}$ comes from condition ($ii$).  For instance, if $H=p^2+V(x)$, where $p$ is momentum, one might try $\sigma_{\sc E}=\psi_{\sc E}(x)/\psi(x)$, which gives $k_{\sc E}=f(E,x)\, p\, \frac{1}{\psi(x)}$ where $f(E,x)$ is a function depending on $E$ and $x$. ($\psi (x) = \langle x|\psi\rangle$ denotes a wavefunction.)  Clearly, $k_E\psi(x)=0$ for all $E$, but $k^\dagger_{\sc E}\phi(x)=0$ implies that $\phi(x)=((f(E,x))^{-1})^*$.  We see that condition ($ii$) cannot be satisfied if we define $\sigma_{\sc E}$ this way. (We can weaken condition ($ii$) so that we require instead $\langle\psi|\phi(E)\rangle>0$ for all $\phi(E)$ with the property $k^\dagger_{\sc E}\phi(E)=0$ and $|\psi\rangle$ as in ($i$)--we will examine  this as well as other issues in a future pulbication.) Furthermore,  this method is quite handy if we know the exact energy eigenstates of the theory. However, there are only a handful of such soluble examples in physics. Most systems can only be solved perturbatively. In this case, the generator $\sigma_{\sc E}$ can only be constructed perturbatively to a given order from a knowledge of the perturbative energy eigenstates to that order. The example that we discuss below, namely, the Lee model \cite{tdlee}, can be solved exactly. However, in a later publication we intend to explore the solubility of $\sigma_{\sc E}$ in greater detail.

\bigskip
\noindent{\bf Example:}

The Lee model \cite{tdlee} corresponds to an interacting field theory describing the decay of  a massive fermion  to another (massive) fermion and a (massive) scalar where all particles are charge neutral.  The running coupling constant, in this  model, becomes imaginary at high energies \cite{kallen} and as a result, this model has been studied recently \cite{leemodel} in the context of ${\cal PT}$-symmetry \cite{bender,mostafazadeh} with an imaginary coupling where the interaction is non-Hermitian.  A quantum mechanical model is described by the Hamiltonian 
\begin{equation}
H= H_{0} + H_{I} = (m_{\theta} \theta^\dagger \theta+m_{\sc V} V^\dagger V+m_{\sc N}N^\dagger N)+ig(\theta^\dagger N^\dagger V+ V^\dagger N \theta),\label{leeH}
\end{equation}
where $V,N$ correspond to the annihilation  operators for the two fermions while $\theta$ represents the annihilation operator for the scalar particle (actually, all particles in this quantum mechanical model are odd under parity, see \cite{leemodel} for the behavior of various operators under parity, ${\cal P}$, and time reversal, ${\cal T}$). The Hamiltonian \eqref{leeH} is invariant under ${\cal PT}$ and satisfies the relation $H={\cal P}H^\dagger {\cal P}$, where ${\cal P} = {\cal P}^{\dagger}$ is the parity operator (${\cal P}^2=\mathbbm{1}$) and in this case, we can identify $S={\cal P}$ which is not positive.  

It is easy to see that the combinations of the number operators $N_{\sc V}+N_{\sc N}$ as well as $N_{\sc V}+N_{\theta}$ commute with the Hamiltonian \eqref{leeH}. As a result, the energy eigenstates can be classified by the quantum numbers $n_{\sc V}+n_{\sc N} = 0,1,2$ (as well as by $n+n_{\sc V} = n, n+1$ where $n$ denotes the eigenvalue of $N_{\theta}$). The states with $n_{\sc V}+n_{\sc N} = 0,2$ are annihilated by the interaction Hamiltonian $H_{I}$ (they correspond to trvial eigenstates of  $H_{0}$) and, therefore, the nontrivial dynamics is contained in the eigenstates with $n_{\sc V}+n_{\sc N}=1$. For any finite $n$, this defines a two dimensional subspace spanned by the following two (unnormalized) eigenstates of $H$ (states are labelled as $|n,n_{\sc V},n_{\sc N}\rangle$):
\begin{eqnarray}
|\Psi_{E_{n}}\rangle &=&\alpha|n,1,0\rangle+\beta|n+1,0,1\rangle = \left(\alpha\ (\theta^{\dagger})^{n}V^\dagger+\beta\ (\theta^{\dagger})^{n+1}N^\dagger\right)|0\rangle,\nonumber\\
|\Psi_{E'_{n}}\rangle &=&(n+1)\bar{\beta}\ |n,1,0\rangle+\bar{\alpha}\ |n+1,0,1\rangle\nonumber\\
& = & \left((n+1)\bar{\beta}\  (\theta^{\dagger})^{n}V^\dagger+\bar{\alpha}\  (\theta^{\dagger})^{ n+1}N^\dagger\right)|0\rangle,\label{leeeigenstates}
\end{eqnarray}
with energy eigenvalues
\begin{eqnarray}
E_{n} & = & \frac{1}{2}\left((2n+1)m_{\theta} +m_N+m_V-\sqrt{\mu^2-4g^2}\right),\nonumber\\
E_{n}' & = & \frac{1}{2}\left((2n+1)m_{\theta} + m_N+m_V+\sqrt{\mu^2-4g^2}\right).
\end{eqnarray}
Here $\alpha =\frac{\mu+\sqrt{\mu^2-4g^2(n+1)}}{2ig}\beta,  \beta =\frac{2g}{\sqrt{\left(\mu+\sqrt{\mu^2-4g^2(n+1)}\right)^2-4g^2(n+1)}}, \mu=(m_\theta+m_{\sc N}-m_{\sc V})$ and $|0\rangle$ denotes the vacuum state of the theory.  We note that for small coupling, the energy eigenvalues are real. We also see from \eqref{leeeigenstates} that the generators for $|\Psi_{E_{n}}\rangle$ and $|\Psi_{E'_{n}}\rangle$ are given by  
\begin{eqnarray}
\sigma_{E_{n}}&=&\alpha\ (\theta^{\dagger})^{n}V^\dagger+\beta\ (\theta^{\dagger})^{n+1}N^\dagger,\nonumber\\
\sigma_{E'_{n}}&=&(n+1)\bar{\beta}\  (\theta^{\dagger})^{n}V^\dagger+\bar{\alpha}\  (\theta^{\dagger})^{n+1}N^\dagger.
\end{eqnarray}
That is, $|\Psi_i\rangle =\sigma_i|0\rangle$, for $i=E_{n},E_{n}'$.  By evaluating $[H,\sigma_i]$ we find $k_i\propto H+ N_{\theta} + N_{\sc N}$, and therefore, $k_i^\dagger|0\rangle=k_i|0\rangle=0$.  Thus, we can choose $|\psi\rangle=|\phi\rangle=|0\rangle$ so that  $q_0=1$.

Let us next solve for $q_{n}$ which represents the action of $q$ restricted to this two dimensional subspace.  In constructing the inverse for $\sigma_{E_{n}}$ we note that there is some freedom following from condition ($iii$), which only requires the inverse to be well defined on $P_{n}$.   We use this freedom to make the additional requirement that $\sigma_{E_{n}}^{-1}|\Psi_{E'_{n}}\rangle =0$.  In this way we avoid constructing a projection onto $|\Psi_{E_{n}}\rangle$.  Doing the same for $\sigma_{E'_{n}}$ we obtain,
\begin{eqnarray}
\sigma_{E_{n}}^{-1}&=&\frac{\bar{\alpha}}{n!}\ \theta^{ n}V-\frac{\bar{\beta}}{n!}\ \theta^{ n+1}N,\nonumber\\
\sigma_{E'_{n}}^{-1}&=&-\frac{\beta}{n!}\ \theta^{ n}V+\frac{\alpha}{(n+1)!}\ \theta^{ n+1}N.
\end{eqnarray}
It can be checked that $\sigma_i^{-1}|\Psi_j\rangle = \delta_{ij} |0\rangle$, as claimed.  Let us look at the product
\begin{eqnarray}
\lefteqn{(\sigma^{-1}_{E_{n}})^\dagger\sigma^{-1}_{E_{n}} = \left(\frac{\alpha}{n!}\ (\theta^{\dagger})^{ n}V^\dagger-\frac{\beta}{n!}\ (\theta^{\dagger})^{ n+1}N^\dagger\right)\left(\frac{\bar{\alpha}}{n!}\ \theta^{ n}V-\frac{\bar{\beta}}{n!}\ \theta^{ n+1}N\right)}\nonumber\\
&=&\frac{\bar\alpha}{n!}\left(\frac{\alpha}{n!}\ (\theta^{\dagger})^{n} V^{\dagger} - \frac{\beta}{n!}\ (\theta^{\dagger})^{n+1} N\right) \theta^{n} V\nonumber\\
&&\qquad -\frac{\bar{\beta}}{n!}\left(\frac{\alpha}{n!}\ (\theta^{\dagger})^{n} V^{\dagger} - \frac{\beta}{n!}\ (\theta^{\dagger})^{n+1} N\right) \theta^{n+1} N\nonumber\\
&=&\frac{1}{n!}\left[|\alpha|^2\  N_{\sc V}(\mathbbm{1}-N_{\sc N})+|\beta|^2\ (n+1) N_{\sc N}(\mathbbm{1}-N_{\sc V})-\alpha\bar{\beta}\ \theta NV^\dagger-\bar{\alpha}\beta \ \theta^\dagger N^\dagger V\right].
\end{eqnarray}
The last relation follows from the fact that the expression acts only on  the states $|n,1,0\rangle$ and $|n+1,0,1\rangle$, where, for instance, we can use $(\theta^{\dagger})^{n}\theta^{n}|n,\cdot,\cdot\rangle = n! |n, \cdot, \cdot\rangle$.  The constraints  $N_{\sc V}+N_{\sc N} = \mathbbm{1}$ (in this space) as well as $N_{\sc V}^{2} = 0 = N_{\sc N}^{2}$, allow us to replace $V^{\dagger}V$ by $N_{\sc V}(\mathbbm{1}-N_{\sc N})$ and $N^{\dagger}N$ by $N_{\sc N} (\mathbbm{1}-N_{\sc V})$.  Likewise, obtain
\begin{eqnarray}
(\sigma^{-1}_{E'_{n}})^\dagger\sigma^{-1}_{E'_{n}}&=&\frac{1}{(n+1)!}\left[|\beta|^2(n+1)\  N_{\sc V}(\mathbbm{1}-N_{\sc N})+|\alpha|^2\  N_{\sc N}(\mathbbm{1}-N_{\sc V})\right.\nonumber\\
& & \qquad\qquad\left. -\alpha\bar{\beta}\ \theta NV^\dagger-\bar{\alpha}\beta \ \theta^\dagger N^\dagger V\right].
\end{eqnarray}

We can sum these two terms, and then make the replacement $n\rightarrow \theta^\dagger \theta = N_{\theta}$ to obtain $q$.  Before doing so, we will exploit the freedom to multiply each $q_{n}$ by an arbitrary positive constant.  Doing this will change the normalization $\langle \Psi_i|\Psi_i\rangle_q=1$, but the innerproduct will remain positive.  Therefore, instead of defining $q_{n}=(\sigma_{E_{n}}^{-1})^\dagger\sigma_{E_{n}}^{-1}+(\sigma_{E'_{n}}^{-1})^\dagger\sigma_{E'_{n}}^{-1}$, we can simplify the expression by multiplying out the factorials dividing the previous two equations (giving instead $\langle \Psi_{E_n}|\Psi_{E_n}\rangle_q=n!$).

With this in mind, we use the identities $|\alpha|^2+(n+1)|\beta|^2=\frac{\mu}{\sqrt{\mu^2-4g^2(n+1)}}$, and $\beta\bar{\alpha}=\frac{ig}{\sqrt{\mu^2-4g^2(n+1)}}$, to obtain 
\begin{eqnarray}
q_{n}&=&n!\;(\sigma_{E_{n}}^{-1})^\dagger\sigma_{E_{n}}^{-1}+(n+1)!\;(\sigma_{E'_{n}}^{-1})^\dagger\sigma_{E'_{n}}^{-1}\nonumber\\
&=&\left(|\alpha|^2+(n+1)|\beta|^2\right)\left(N_{\sc V}(\mathbbm{1}-N_{\sc N})+N_{\sc N}(\mathbbm{1}-N_{\sc V})\right)-2\alpha\bar{\beta}\ \theta NV^\dagger-2\beta\bar{\alpha}\ \theta^\dagger N^\dagger V\nonumber\\
&=&\frac{\mu}{\sqrt{\mu^2-4g^2(n+1)}}\left(N_{\sc V}(\mathbbm{1}-N_{\sc N})+N_{\sc N}(\mathbbm{1}-N_{\sc V})\right)\nonumber\\
&&+\frac{2ig}{\sqrt{\mu^2-4g^2(n+1)}}\ \theta NV^\dagger-\frac{2ig}{\sqrt{\mu^2-4g^2(n+1)}}\ \theta^\dagger N^\dagger V.
\end{eqnarray}
In order to make the expression valid for an arbitrary number of scalars we replace $n$ by $\theta^\dagger \theta =N_\theta$.  We must be careful about the operator ordering, for instance, $\theta^\dagger$ increases the scalar particle number by one.  In addition, the fermion operators $V,N$ annihilate the groundstate so that we should add a projection onto $|0\rangle$ given by $P_{0}= (\mathbbm{1}-N_{\sc N})(\mathbbm{1}-N_{\sc V})$ to finally get:
\begin{eqnarray}
q&=&P_0+\sum_{n}q_{n}\stackrel{}{\longrightarrow}(n\rightarrow N_\theta)\nonumber\\
&=&\mathbbm{1}-N_{\sc N}-N_{\sc V}+N_{\sc N}N_{\sc V}+\frac{\mu N_{\sc V}(\mathbbm{1}-N_{\sc N})}{\sqrt{\mu^2-4g^2(N_\theta+1)}}+\frac{\mu N_{\sc N}(\mathbbm{1}-N_{\sc V})}{\sqrt{\mu^2-4g^2 N_\theta}}\nonumber\\
&&+\theta NV^\dagger\ \frac{2ig}{\sqrt{\mu^2-4g^2N_\theta}}-\frac{2ig}{\sqrt{\mu^2-4g^2N_\theta}}\ \theta^\dagger N^\dagger V.
\end{eqnarray}
The above expression was previously determined \cite{leemodel} using a more elaborate method.  In that case, the result was derived, in part, by requiring that the operator ${\cal C}\equiv q{\cal P}$ is an involution (${\cal C}^2=\mathbbm{1}$) when acting on the eigenstates of $H$ (it is actually a projection onto the set $\{|\Psi_{E_{n}}\rangle,\:|\Psi_{E_{n}'}\rangle\}$).  

In summary, we have utilized the fact that the generators of eigenstates of a non-Hermitian operator, $H$, contain sufficient information for constructing a positive inner product space $\mathcal{H'}$. We have illustrated how the method works in the case of the ${\cal PT}$ symmetric Lee model \cite{leemodel}. In a subsequent publication we plan to consider possible constructions for the operator $\sigma_E$, as well as its role in more sophisticated models.

\bigskip

\noindent{\bf Acknowledgments}

One of us (A. D.) would like to thank Prof. Carl Bender for many helpful discussions on ${\cal PT}$ symmetry. He would also like to thank Prof. Jihn E. Kim and the members of the theoretical physics group of the Seoul National University for hospitality where part of this work was done. This work was supported in part  by US DOE Grant number DE-FG 02-91ER40685.

\end{document}